\documentclass[fleqn,10pt]{wlscirep}
\usepackage[utf8]{inputenc}
\usepackage[T1]{fontenc}
\usepackage{graphicx}
\usepackage{multirow}
\usepackage{array}
\usepackage{multirow,amssymb,amsbsy,amsmath,epstopdf}
\usepackage{color}
\usepackage{ulem}
\usepackage{subfigure}
\title{Activation of quantum steering sharing with unsharp nonlocal product measurements}

\author[1,2]{Xin-Hong Han}
\author[1]{Tian Qian}
\author[1]{Shan-Chuan Dong}
\author[1+]{Ya Xiao}
\author[1,*]{Yong-Jian Gu}
\affil[1]{College of Physics and Optoelectronic Engineering, Ocean University of China, Qingdao 266100, People's Republic of China}

\affil[2]{College of Computer Science and Technology, Shandong University of Technology, Zibo, 255000, People's Republic of China}
\affil[+]{xiaoya@ouc.edu.cn}
\affil[*]{yjgu@ouc.edu.cn}

\begin{abstract}
	
Quantum steering is commonly shared among multiple observers by utilizing unsharp measurements. However, their usage is limited to local measurements and is not suitable for nonlocal-measurement-based cases. Here, we present a novel approach in this study, suggesting a highly efficient technique to construct optimal nonlocal measurements by utilizing quantum ellipsoids to share quantum steering. This technique is suitable for any bipartite state and offers benefits even in scenarios with a high number of measurement settings. Using the Greenberger-Horne-Zeilinger state as an illustration, we show that employing unsharp nonlocal product measurements can activate the phenomenon of steering sharing in contrast to using local measurements. Moreover, our findings demonstrate that nonlocal measurements with unequal strength possess a greater activation capability compared to those with equal strength. Our activation method differs from previous ones as it eliminates the need to copy the shared states or diminish other quantum correlations, thus making it convenient for practical experimentation and conservation of resources.

\end{abstract}

\begin{document}

\flushbottom
\maketitle
\thispagestyle{empty}

\section*{Introduction}

In 1936, Schr\"odinger first proposed the concept of quantum steering as a response to the EPR paradox \cite{Schrodinge1936,einstein1935can}. Many years later, Wiseman et al. strictly redefined quantum steering using a local hidden variable and local hidden state (LHV-LHS) model \cite{wiseman2007steering}. It sits between Bell nonlocality \cite{bell1964einstein} and quantum entanglement \cite{einstein1935can} and exhibits a distinctive asymmetric property \cite{handchen2012observation,he2013genuine,bowles2014one,gallego2015resource,cavalcanti2015detection,sun2016experimental,xiao2017demonstration,uola2020quantum,marton2021cyclic}. As an essential type of quantum resource, quantum steering has great applications in quantum key distribution \cite{branciard2012one,walk2016experimental}, subchannel discrimination \cite{sun2018demonstration}, asymmetric quantum network \cite{cavalcanti2015detection}, randomness generation \cite{skrzypczyk2018maximal,guo2019experimental} and randomness certification \cite{curchod2017unbounded}. Improving the utilization efficiency of quantum steering is of great importance, not only for fundamental quantum information science but also for applications in quantum communication. Several methods have been developed, including activation of quantum steering and sharing of quantum steering.

One method to activate steering of a quantum state is to perform local collective measurements on multicopy of the state \cite{quintino2016superactivation,hsieh2016quantum,pan2019activating}. However, this requires multicopy if the dimension of the state is low, or a high-dimensional state if the copy number is small, which is impractical \cite{quintino2016superactivation}. An alternative method is to use local filtering on a single copy of the quantum state, which employs non-unitary operators to convert a full rank state into a normal form (such as a maximally entangled state) by weakening other types of correlations in the quantum system \cite{pramanik2019revealing}.

To overcome the drawbacks of the aforementioned methods, researchers have relaxed the no-signaling condition and found that the steering of a single copy of the entangled states can be shared among multiple sequential observers either by unsharp measurements \cite{sasmal2018steering,shenoy2019unbounded,choi2020demonstration,han2021sharing,gupta2021genuine,zhu2022einstein,han2022manipulating} or standard projective measurements \cite{steffinlongo2022projective,xiao2022experimental}.
This method, known as steering sharing, has been extensively studied in bipartite systems and has also been used to investigate the reuse of genuine multipartite steering and network steering.

Until now, all research aimed at improving the utilization efficiency of quantum steering has been restricted to local measurements, which are not suitable for some quantum information tasks that require nonlocal measurements, such as quantum teleportation. Indeed, nonlocal measurements between spatially separated observers cannot be accomplished through local measurements and classical communication. 
This raises some interesting questions: Is it possible to transform an unsteerable state into a steerable one by employing unsharp nonlocal measurements? If yes, how can we construct these optimal nonlocal measurements? In particular, can unsharp nonlocal measurements still be more effective than unsharp local measurements in activating steering, even though each observer obtains the same classical measurement outcomes?

In this paper, we propose a steering sharing scenario using sequential unsharp nonlocal measurements. Focusing on the linear steering criterion for $n$-setting measurements, we develop a method to efficiently construct these optimal nonlocal measurements with the help of steering ellipsoids. As an example, we consider the Greenberger-Horne-Zeilinger (GHZ) state in the case of two-setting measurements, and demonstrate that using unsharp nonlocal product measurements can activate more steering sharing than using unsharp local measurements. We quantify the measurement strength ranges that can be used to activate the steering sharing and find that these ranges can be further extended by replacing equal-strength nonlocal measurements with unequal-strength ones. Our activation method only requires a single copy of the state and does not weaken other types of correlations in the quantum system, making it more experiment-friendly and resource-reusable. 

The paper is organized as follows: In Section II, we present the steering sharing scenario based on sequential unsharp nonlocal measurements. The method used to construct the optimal nonlocal measurement settings is described in Section III. An illustration of steering sharing activation is given in Section IV. Finally, we present the conclusion and some outlooks in Section V.

\section*{Steering sharing using sequential unsharp nonlocal measurements}
%%%%%%%%%%%%%%%%%%%%%%%%%%%%%%%%%%%%%%%%%%%%%%%%%%%%%%%%%%
%Fig.1
%%%%%%%%%%%%%%%%%%%%%%%%%%%%%%%%%%%%%%%%%%%%%%%%%%%%%%%%%%
\begin{figure}[!ht]
	\centering\includegraphics[width=6cm]{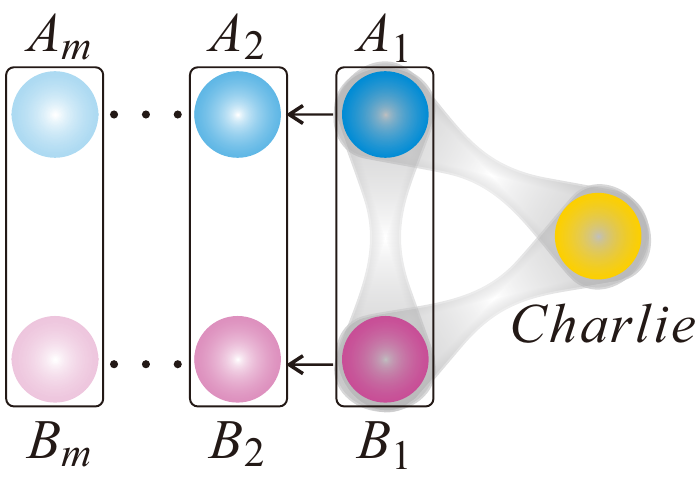}
	\caption{The scenario of steering sharing using sequential unsharp nonlocal measurements. A three-qubit state is initially shared between the spatially separated $A_1$, $B_1$ and Charlie. $A_1$ and $B_1$ perform nonlocal measurement on their qubits and transmit the post-measurement qubits to $A_2$ and $B_2$. This process is repeated until the last pair of $A_m$ and $B_m$ measure the qubits. Meanwhile, Charlie performs local measurements on his single qubit. The goal is for each pair of $A_i$ and $B_i$ to remotely steer the quantum state of Charlie simultaneously and independently.} 
	\label{FIG_1}
\end{figure}
%%%%%%%%%%%%%%%%%%%%%%%%%%%%%%%%%%%%%%%%%%%%%%%%%%%%%%%%%%%%%%%%%%%%%%%%
Figure~\ref{FIG_1} illustrates a steering sharing scenario based on sequential unsharp nonlocal measurements. A three-qubit state $\rho_{ABC} $ is shared among Charlie and multiple pairs of observers, labeled as $A_i$ and $B_i$ where $i\in\{1,2,...,m\}$. The task of each pair of $A_i$ and $B_i$ is to remotely steer the quantum state of Charlie, simultaneously and independently. Charlie will be convinced by $A_i$ and $B_i$ if the correlation between their measurement outcomes cannot be explained by the LHV-LHS model. In our scenario, each pair of $A_i$ and $B_i$ (except the last $ A_{m}$ and $B_{m}$) must perform unsharp measurements to generate strong correlations between them and Charlie to elude the LHV-LHS model, while preserving enough entanglement for the next pair of observers to achieve the same goal. Suppose the $k$-th nonlocal measurement setting of $A_i$ and $B_i$ is $\hat{\Pi}_{k}^{(i)}$ and the corresponding measurement strength is $\lambda_{k}^{(i)}$. And the $k$-th local sharp measurement setting of Charlie is denoted as $\hat{\Lambda}_k$. The success of the steering task among $ A_{i}$, $ B_{i}$ and Charlie can be tested by violating the bipartite $ n $-setting linear steering inequality of the form \cite{saunders2010experimental}
\begin{equation}
	S_n^{(i)}\!\equiv\!\frac{1}{n}\sum_{k=1}^n \lambda_{k}^{(i)}\langle  \hat{\Pi}_{k}^{(i)}\otimes\hat{\Lambda}_{k}\rangle\!\leq C_{n}. 
	\label{Sn}
\end{equation}
The bound $C_{n}=\sum_{\{\Lambda_{k}\}}\lambda_{max}(\frac{1}{n}\sum_{k=1}^n \langle  \hat{\Pi}_{k}^{(i)} \Lambda_{k}\rangle)$ is the maximum value of steering parameter $ S_n^{(i)} $ if LHV-LHS model exists, where $ \Lambda_{k}\in\{-1,1\} $ denotes a random variable and  $\lambda_{max}(\hat{E} )$ denotes the largest eigenvalue of $\hat{E} $.
The expectation value $ \langle\hat{\Pi}_{k}^{(i)}\otimes\hat{\Lambda}_{k}\rangle={\rm Tr}[\hat{\Pi}_{k}^{(i)}\otimes\hat{\Lambda}_{k} \rho_{ABC}^{(i)} ]$ is evaluated with respect to the average post-measurement states $ \rho_{ABC}^{(i)}$ shared among $A_{i}$, $B_{i}$, and Charlie, which can be expressed as  
\begin{equation}
	\rho_{ABC}^{(i)}=\frac{1}{n}\!\sum_{k=1}^{n} \sum_{o_{k}\!=\!-1,1}[(\!K^{(i)}_{o_{k}}\otimes I_C )\rho_{ABC}^{(i-1)}({K^{(i)}_{o_{k}}}\!^{\dagger}\otimes I_C)], 
\end{equation}
where $K^{(i)}_{o_{k}}{K^{(i)}_{o_{k}}}\!^{\dagger}=\hat{\Pi}_{k}^{(i)}$ and $o_{k}$ indicates the outcome of $A_{i}$ and $B_{i}$, and $o_{k}\in\{-1,1\}$.

To demonstrate the existence of steering ability among $A_{i}$, $B_{i}$ and Charlie, it is essential to ensure that their measurement settings $\{\hat{\Pi}_{k}^{(i)}, \hat{\Lambda}_{k}\}$ can achieve the best violations of the linear steering inequality Eq.~(\ref{Sn}). The previous method \cite{zheng2017optimized} is to first fix the measurement setting $\hat\Lambda_{k}$ in a certain direction, and then explore the corresponding measurement setting $\hat{\Pi}_{k}^{(i)}$ to maximize $\langle \hat{\Pi}_{k}^{(i)}\otimes\hat{\Lambda}_{k}\rangle$; then, change $\hat\Lambda_{k}$ to another direction and repeat the above process; finally, by searching $\hat{\Lambda}_{k}$ in the entire operator space, the measurement settings  $\{\hat{\Pi}_{k}^{(i)}, \hat\Lambda_{k}\}$ that maximize the difference between $S_n^{(i)}$ and $C_n$ can be obtained. This is computationally expensive for arbitrary states, especially when the number of settings or the system dimension is large. We aim to develop a practical scheme to determine the optimal measurement settings $\{\hat{\Pi}_{k}^{(i)}, \hat{\Lambda}_{k}\}$ for detecting steering in the scenario depicted in Figure~\ref{FIG_1}.

\section*{Methods for finding optimial measurement settings}
Note that the more optimal the measurement settings of the steering party are, the closer the corresponding conditional states of the steered party are to the surface of the Bloch sphere. Quantum steering ellipsoid represents the states that the steering party can collapse the steered party to, considering all possible measurements performed on his subsystem \cite{jevtic2014quantum}. Therefore, the optimal measurement setting can be easily determined by analyzing the geometric characteristics of the corresponding steering ellipsoid. However, since it is only available for two-qubit systems, we need to compress the three-qubit state into a two-qubit state. Here, we consider the case where the reduced state of $A_{i}$ and $B_{i}$ can be represented using only two of the four computational bases, namely $\{|00\rangle, |01\rangle, |10\rangle,|11\rangle \}$. If we treat one basis as $\vert\widetilde 0\rangle$ and the other as $\vert\widetilde 1\rangle$, then the state $\rho_{ABC}^{(i)}$ shared among $A_{i}$, $B_{i}$, and Charlie can be compressed as a ``two-qubit" state, which can be written in the following form:  
\begin{equation}
	\rho_{ABC}^{(i)}=\frac{1}{4}( \widetilde I\otimes I  +\widetilde m \cdot \widetilde\sigma \otimes I + \widetilde I\otimes \vec{n}\cdot\vec{\sigma} + \sum_{\mu,\nu=1}^3 T_{\mu,\nu} \widetilde\sigma_{\mu}\otimes\sigma_{\nu}.
\end{equation}
where $\widetilde I $ and $I$ are identity operator, $\widetilde {m}$ and $\vec{n}$ are the Bloch vectors  of the reduced states $\rho_{AB}^{(i)}$ and $\rho_{C}$ of $\rho_{ABC}^{(i)}$, $\widetilde{\sigma}\!\equiv\!\{\widetilde\sigma_{x},\widetilde\sigma_{y},\widetilde\sigma_{z}\}$ and $\vec{\sigma}\!\equiv\!\{\sigma_{x},\sigma_{y},\sigma_{z}\}$ are the vectors of Pauli spin operators,  $T_{\mu,\nu}$ is the element of correlation matrix $T$. 

Considering all possible measurements by $A_i$ and $B_i$, the state of Charlie can be steered to an ellipsoid $\Omega _{C}^{(i)} $, which is centered at $o_{C}^{(i)} = (\vec{n} - T\widetilde {m})/(1-|{\widetilde{m}}|^2)$. The orientation and the squared lengths of the ellipsoid's semiaxes are given by the eigenvectors and eigenvalues of the ellipsoid matrix \cite{jevtic2014quantum} 
\begin{equation}
	O_{C}^{(i)}=\dfrac{(T-\vec{n}\widetilde{m}^{\intercal})}{1-|{\widetilde{m}}|^2}(I+\dfrac{\widetilde{m}\widetilde{m}^{\intercal}}{1-|{\widetilde{m}}|^2 })(T^{\intercal}-\widetilde{m}\vec{n}^{\intercal}).
\end{equation}
Similarly, the steering ellipsoid $\Omega_{AB}^{(i)}$ of $A_i$ and $B_i$ can be obtained by swapping the roles of Charlie with $A_i$ and $B_i$. 

With the steering ellipsoid, one can easily determine which states are closer to the surface of the Bloch sphere, and thus obtain the corresponding optimal measurement settings. For example, the optimal two- or three-setting measurement directions are typically aligned with the principal axes of the steering ellipsoid. If the optimal measurement setting for Charlie is fixed, such as chosen as  $\hat{\Lambda}_k$, one can straightforwardly calculate the corresponding optimal measurement setting $\hat{\Pi}_{k}^{(i)}$ for $A_{i}$ and $B_{i}$  with the following two conditions. Firstly, the eigenvector list $\{e_{k}^{(i)}\}$ of $\hat{\Pi}_{k}^{(i)}$  is the same as the normalized conditional state $\widetilde\rho_{AB}^{(i)}$  of $A_{i}$ and $B_{i}$ after Charlie measures his qubit by $\hat{\Lambda}_k$. Secondly, the eigenvector list $\{\alpha_{k}^{(i)}\}$ of $\hat{\Pi}_{k}^{(i)}$ has the same order with the eigenvalue list $\{\beta_{k}^{(i)}\}$ of $\widetilde\rho_{AB}^{(i)}$~\cite{zheng2017optimized}. And the situation is similar when the measurement setting $\hat{\Pi}_{k}^{(i)}$ of $A_{i}$ and $B_{i}$ is fixed. Clearly, the method proposed here to find optimal measurement settings can quickly reduce the computational complexity.
%%%%%%%%%%%%%%%%%%%%%%%%%%%%%%%%%%%%%%%%%%%%%%%%%%%%%%%%%%%%%%%%%%%%%
%%%%%%%%%%%%%%%%%%%%%%%%%%%%%%%%%%%%%%%%%%%%%%%%%%%%%%%%%%
%Fig.2
%%%%%%%%%%%%%%%%%%%%%%%%%%%%%%%%%%%%%%%%%%%%%%%%%%%%%%%%%%
\begin{figure}[!ht]
	\centering
	\subfigure[]{\includegraphics[width=6cm]{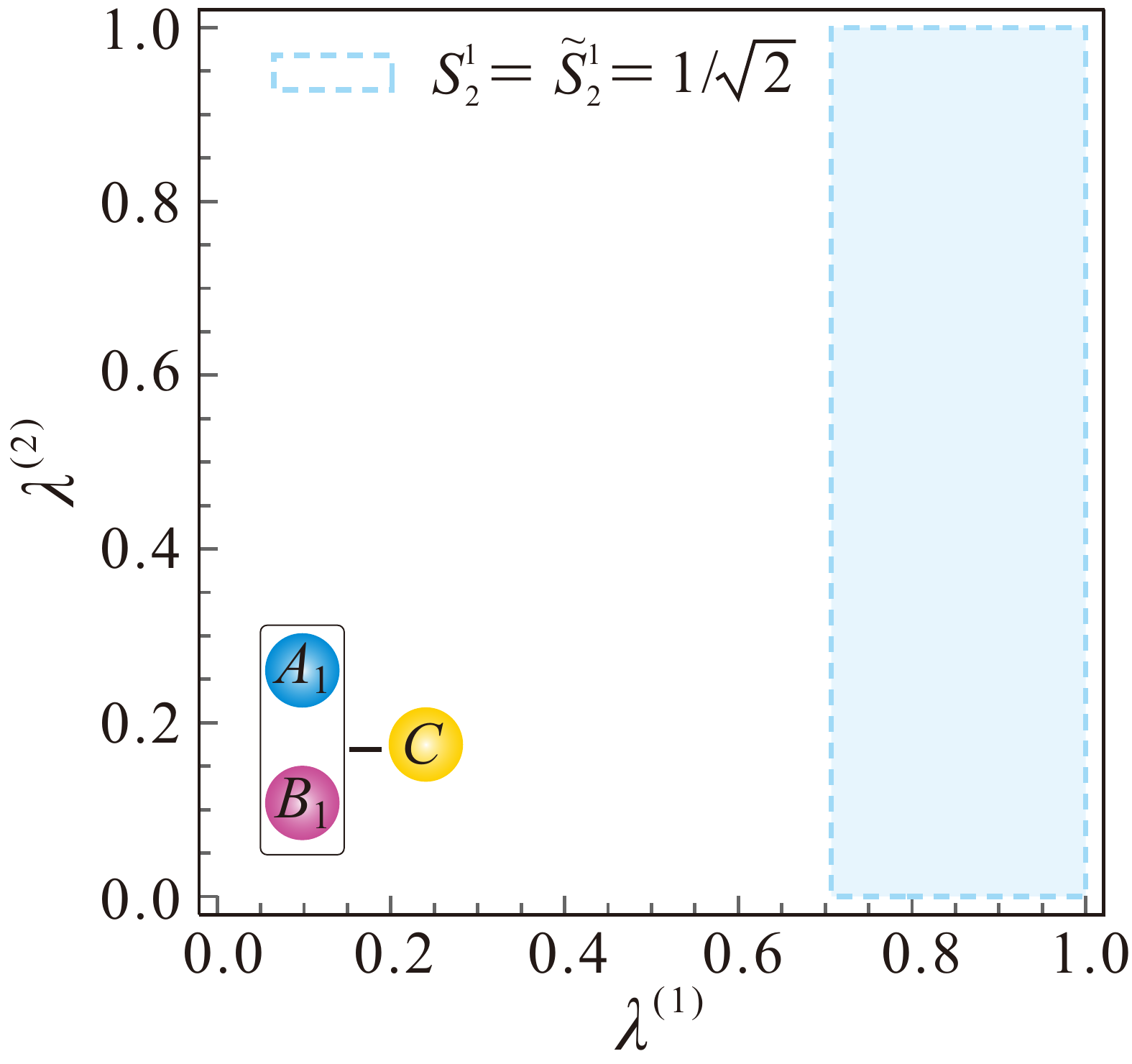}}
	\subfigure[]{\includegraphics[width=6cm]{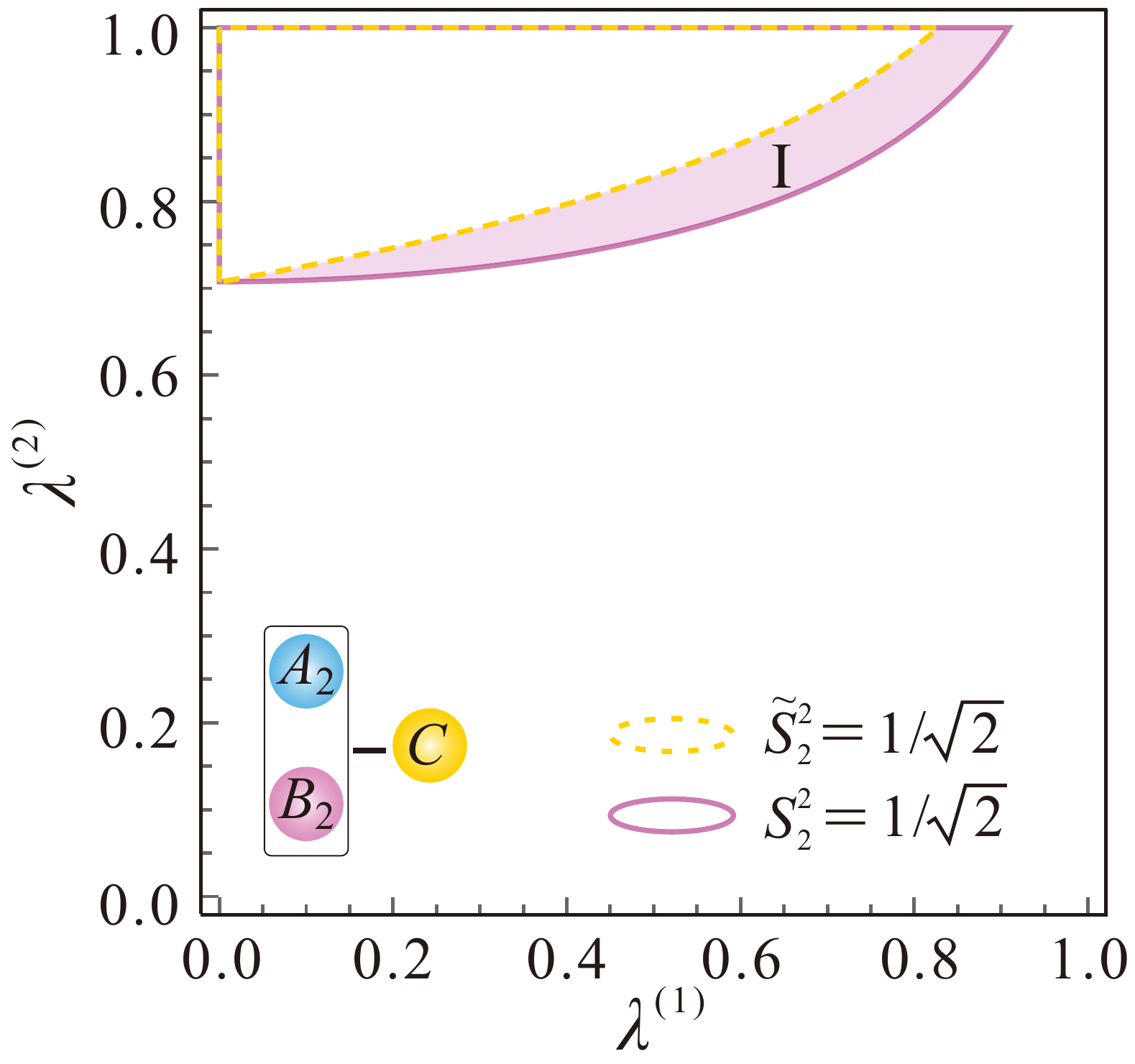}}
	\subfigure[]{\includegraphics[width=6cm]{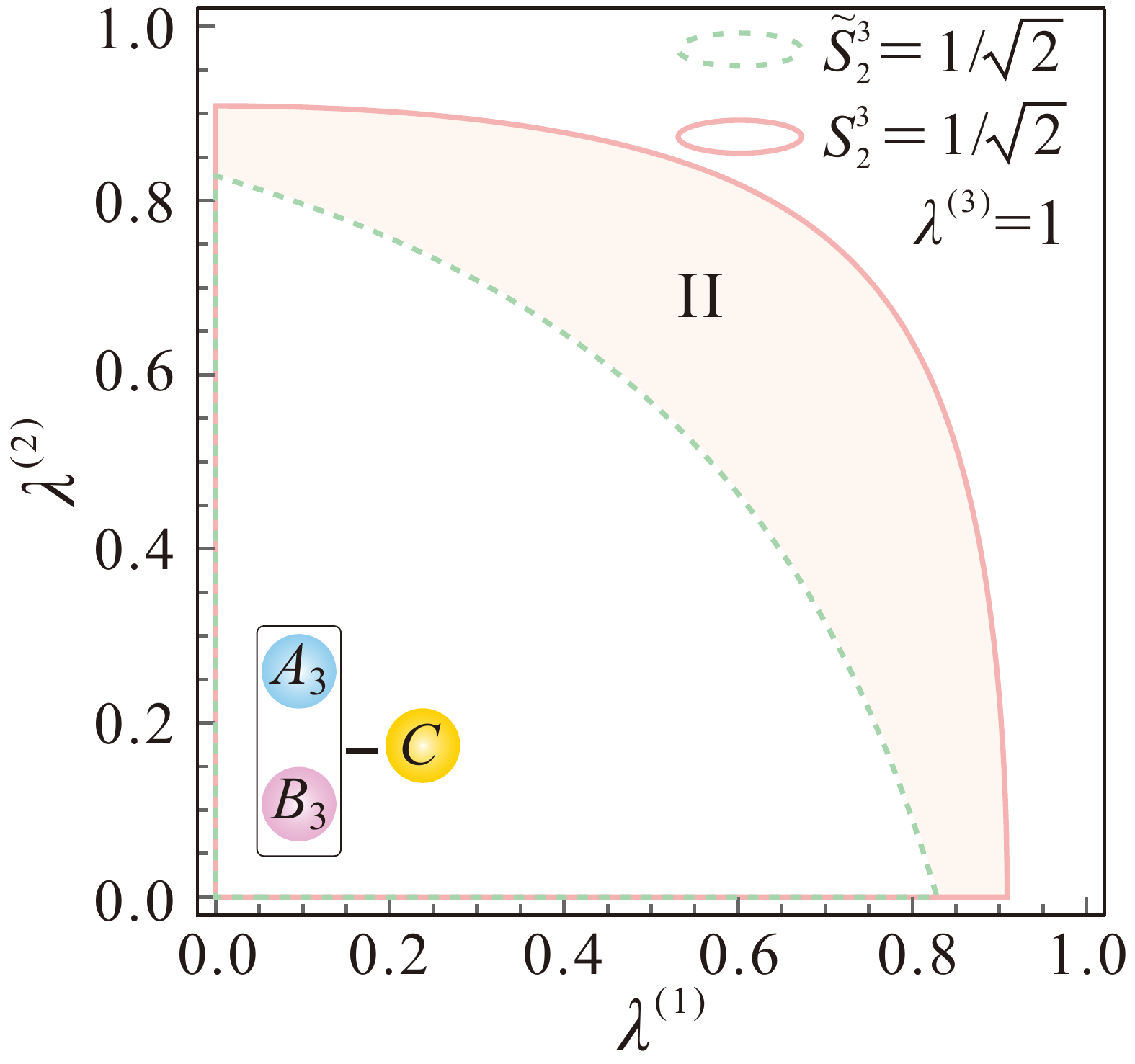}}
	\subfigure[]{\includegraphics[width=6cm]{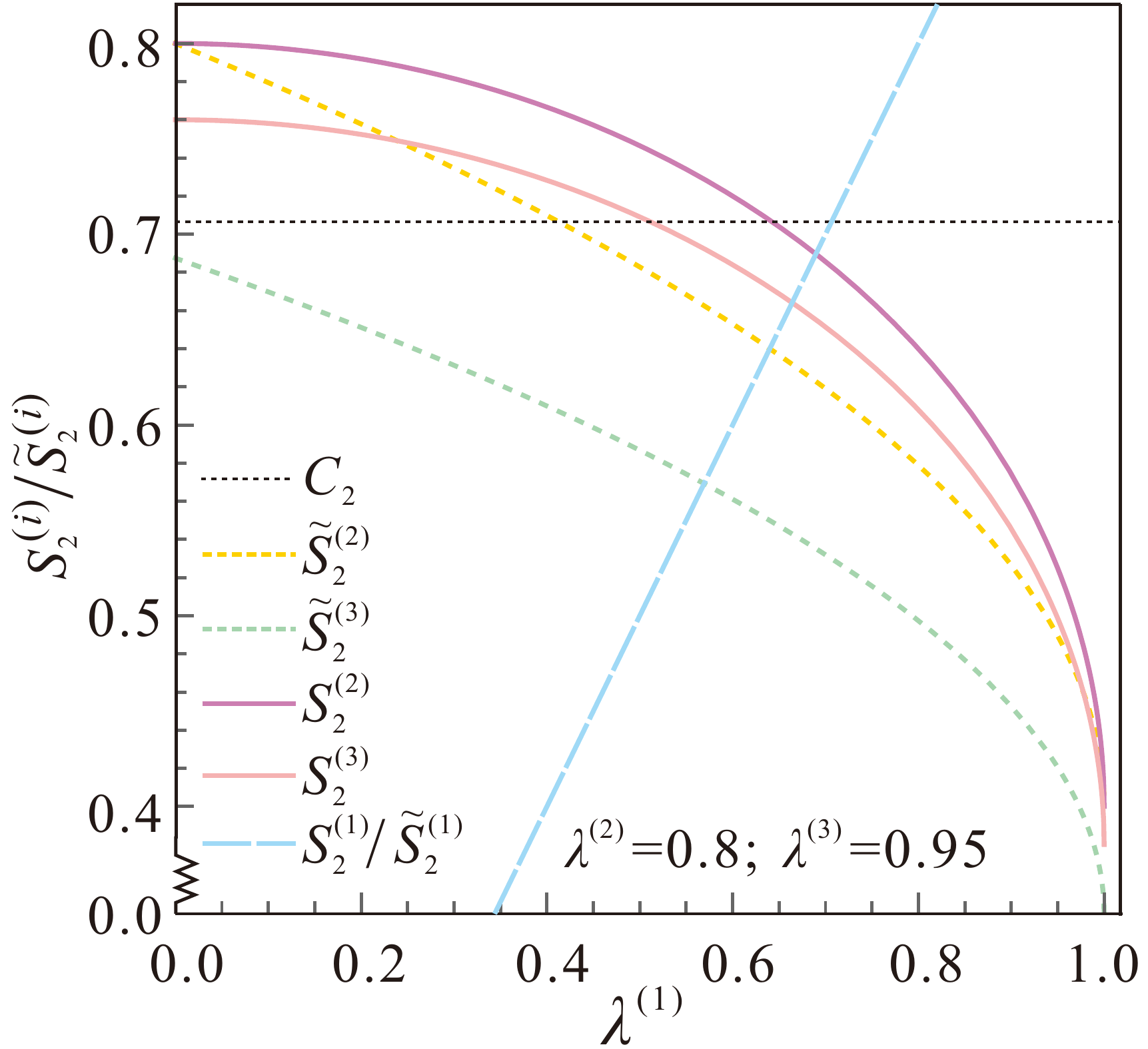}}
	\caption{The comparison diagram illustrating the successful steering regions for unsharp nonlocal measurements versus unsharp local measurements. Specifically, (a), (b), and (c) respectively show the successful steering regions for the first, second, and third pairs of $A_i$ and $B_i$. In each of these diagrams, the solid and dotted curves represent the boundaries of $S_2^{(i)}=1/\sqrt{2}$ and $\widetilde S_2^{(i)}=1/\sqrt{2}$, respectively. The regions within these boundaries indicate successful steering between the corresponding pairs of observers. Notably, the purple region (labeled as I) and the pink region (labeled as II) correspond to ${S_{2}^{(2)}\!>\!1\sqrt{2}\&\widetilde S_{2}^{(2)}\leq 1\sqrt{2}}$ and ${S_{2}^{(3)}\!>\!1\sqrt{2}\&\widetilde S_{2}^{(3)}\leq 1\sqrt{2}}$ respectively. (d) Steering parameters $S_2^{(i)}$ and $\widetilde S_2^{(i)}$ as a function of the measurement strength $\lambda^{(1)}$ when $\lambda^{(2)}=0.8$ and $\lambda^{(3)}=0.95$.}
	\label{FIG_2}
\end{figure}
\section*{Example of steering sharing activation}
 In this section, we use the three-qubit GHZ state $\vert\mathrm{GHZ}\rangle =(\vert 000\rangle+\vert111\rangle)/\sqrt{2}$ as an illustration to show how sequential unsharp nonlocal product measurements can activate the sharing ability of steering that cannot be realized by sequential unsharp local measurements. To simplify the analysis, we focus on the case of two-setting measurements. By defining $\vert 00\rangle\equiv\vert\widetilde 0 \rangle$ and $\vert 11\rangle\equiv\vert\widetilde 1\rangle$, we can compress the GHZ state in terms of a two-qubit state.
 Using the method shown in the previous section, we can obtain the steering ellipsoid $\Omega_{AB}^{(i)}$ of $A_i$ and $B_i$, as well as the steering ellipsoid $\Omega_{C}^{(i)}$ of Charlie. It is evident that the principal axes of these steering ellipsoids are $\vec{x}$, $\vec{y}$  and $\vec{z}$. Taking measurement strength into consideration, the optimal two-setting  measurements can be set as $\{\hat{\Pi}_{1}^{(i)}=\lambda_{1}^{(i)}(\sigma_y\otimes\sigma_y), \hat{\Lambda}_{1}=\sigma_x\}$ and $\{\hat{\Pi}_{2}^{(i)}=\lambda_{2}^{(i)}(\sigma_y\otimes\sigma_x), \hat{\Lambda}_{2}=\sigma_y\}$. And the steering inequality Eq.~(\ref{Sn}) can be rewritten as
 \begin{equation}
 	S_2^{(i)}\!=\!\frac{1}{2^i}\Big{[}\lambda_{2}^{(i)}\!\prod\limits_{1\leq j\leq i\!-\!1}\!(1\!+\!F_{\lambda_{1}^{(j)}})+\lambda_{1}^{(i)}\!\prod\limits_{1\leq j\leq i\!-\!1}\!(1\!+\!F_{\lambda_{2}^{(j)}})\Big{]}\!\leq\!C_{2},
 	\label{Eq.5}
 \end{equation}
 where $F_{\lambda_{1}^{(j)}}=\sqrt{1-({\lambda_{1}^{(j)}})^2}$, $F_{\lambda_{2}^{(j)}}=\sqrt{1-({\lambda_{2}^{(j)}})^2}$ and $C_{2}=1/\sqrt{2}$.
 
 In comparison, when unsharp local measurements are used, the measurement settings for $A_i$, $B_i$, and Charlie would be $\{\hat{M}_{1}^{A_i}=\eta_{1}^{(i)} \sigma_y, \hat{M}_{1}^{B_i}= \gamma_{1}^{(i)} \sigma_y, \hat{M}_{1}^{C}=\sigma_x\}$  and $\{\hat{M}_{2}^{A_i}=\eta_{2}^{(i)}\sigma_y,\hat{M}_{2}^{B_i}=\gamma_{2}^{(i)}\sigma_x, \hat{M}_{2}^{C}=\sigma_y\}$. The corresponding steering parameter $ \widetilde S _2^{(i)} $ can be expressed as
 \begin{equation}
 	\widetilde S _2^{(i)}\!=\!\frac{1}{2^{i}}\Big{[}\lambda_{2}^{(i)}\!\prod\limits_{1\leq j\leq i\!-\!1}\!(1\!+\!F_{\gamma_{1}^{(j)}})+\lambda_{1}^{(i)}\!\prod\limits_{1\leq j\leq i\!-\!1}\!(1\!+\!F_{\gamma_{2}^{(j)}})\Big{]}\leq\!C_{2},
 	\label{Eq.6}
 \end{equation}
 where $F_{\gamma_{1}^{(j)}}=\sqrt{1-({\gamma_{1}^{(j)}})^2}$ and $F_{\gamma_{2}^{(j)}}=\sqrt{1-({\gamma_{2}^{(j)}})^2}$.
 The specific values of measurement strength parameters $ \lambda_{k}^{(i)} $, $\eta_{k}^{(i)}$ and $\gamma_{k}^{(i)}$ should satisfy $\lambda_{k}^{(i)}=\eta_{k}^{(i)}*\gamma_{k}^{(i)}$, $ k\in \{1,2\} $. It should be noted that since the GHZ state is an eigenstate of  $ \hat{M}_{k}^{A_i}\otimes\hat{M}_{k}^{B_i}\otimes\hat{M}_{k}^{C} $, these measurement settings are still optimal in the case of unsharp local measurements.
\subsection*{The measurement settings with equal strength} 
We first investigate the activation of steering sharing when the strength of the two-setting nonlocal measurements used by $A_i$ and $B_i$ is equal, i.e., $\lambda^{(i)}=\lambda_{1}^{(i)}=\lambda_{2}^{(i)}$, and the strength of two-setting local measurements used by each observer is equal, i.e., $ \eta^{(i)}=\eta_{1}^{(i)}=\eta_{2}^{(i)} $, $\gamma^{(i)}\!=\!\gamma_{1}^{(i)}\!=\!\gamma_{2}^{(i)}$. The steering parameters in Eq.~(\ref{Eq.5}) and Eq.~(\ref{Eq.6}) can be respectively rewritten as $ S_2^{(i)}=\frac{1}{2^{i-1}}\Big{[}\lambda^{(i)}\prod\limits_{1\leq j\leq i\!-\!1}(1\!+\!\sqrt{1-({\lambda^{(j)}})^2})\Big{]}$ and $\widetilde S _2^{(i)}=\frac{1}{2^{i-1}}\Big{[}\lambda^{(i)}\prod\limits_{1\leq j\leq i\!-\!1}(1\!+\!\sqrt{1-({\gamma^{(j)}})^2})\Big{]}$. Obviously, in order to satisfy the condition of  $\lambda^{(i)}=\eta^{(i)}*\gamma^{(i)}$, if $\lambda^{(i)}<\gamma^{(i)}$, then $ S _2^{(i)}>\widetilde S _2^{(i)}$. In other words, more steering sharing can be discovered in the scenario of nonlocal product measurements compared to that of local measurements. As a result, unsharp nonlocal product measurement can be used to activate the sharing ability of steering. 

To clarify the effects of nonlocal product measurements on activating steering sharing, we set  $\eta^{(i)} = \gamma^{(i)} = \sqrt{\lambda^{(i)}}$, where $i \in \{1, 2, 3\}$. Figure~\ref{FIG_2} (a), (b), and (c) show the steering regions for the first, second, and third pairs of $A_{i}$ and $B_{i}$, respectively. These regions are parameterized by the measurement strength $\lambda^{(1)}$ and $\lambda^{(2)}$. Obviously, the ranges of $\lambda^{(1)}$ and $\lambda^{(2)}$ that can be used to verify the existence of steering among $A_{1}$, $B_{1}$ and Charlie are the same in both unsharp nonlocal and local measurements. However, as the number of the pairs of $A_{i}$ and $B_{i}$ increases, the ranges of $\lambda^{(1)}$ and $\lambda^{(2)}$ that satisfy $ S _2^{i}>1\sqrt{2} $ are larger than the ranges that satisfy $\widetilde S _2^{i}>1\sqrt{2} $. 
For example, in Figure~\ref{FIG_2} (b) and (c), the steering regions I (marked in purple) and II (marked in pink) can be activated with unsharp nonlocal measurements.
This suggests that nonlocal product measurement can activate steering sharing that is impossible to achieve through local measurement. 

In Figure~\ref{FIG_2}(d), we present the steering parameters $S_2^{(i)}$ and $\widetilde S_2^{(i)}$ ($i\in\{1,2,3\}$) varying with the measurement strength $\lambda^{(1)}$.
$S_2^{(1)}$ and $\widetilde S_2^{(1)}$ are represented by the same dashed blue line. $S_2^{(2)}$ and $\widetilde S_2^{(2)}$ are represented by the solid purple and dotted yellow lines, respectively. $S_2^{(3)}$ and $\widetilde S_2^{(3)}$ are represented by the solid pink and dotted green lines, respectively. Especially, when the measurement strength $\lambda^{(1)}=0.5$,
$S_2^{(2)}=0.75$ and is greater than $1\sqrt{2}$, while $\widetilde S_2^{(2)}=0.68$ and is less than $1\sqrt{2}$, Therefore, unsharp nonlocal measurements can activate more steering. When $\lambda^{(1)}=0.4$, the amounts by which $S _2^{(2)}$ and $\widetilde S_2^{(2)}$ exceed the classical bound $1\sqrt{2}$ are 0.003 and 0.059, respectively. This means that steering can be easier to implement in experiments via unsharp nonlocal measurements. Additionally, it is clear that the values of $\widetilde S_2^{(3)}$ are consistently lower than the classical bound $C_2$, whereas $S_2^{(3)}$ has the potential to exceed $C_2$. And it is possible for $S_2^{(3)}$ to exceed $\widetilde S_2^{(2)}$. For example, when $\lambda_{1}=0.4$, $\widetilde S_2^{(2)}$ and $\widetilde S_2^{(3)}$ have values of 0.71 and 0.61, respectively, while $S_2^{(2)}$ and $S_2^{(3)}$ have values of 0.77 and 0.73, respectively. It is also evident that there are no values of $\lambda^{(1)}$ that satisfy $S_{2}^{(1)}>1\sqrt{2}$, $S_{2}^{(2)}>1\sqrt{2}$, and $S_{2}^{(3)}>1\sqrt{2}$. This indicates that at most two pairs of $A_{i}$ and $B_{i}$ can share steering with Charlie simultaneously.
%%%%%%%%%%%%%%%%%%%%%%%%%%%%%%%%%%%%%%%%%%%%%%%%%%%%%%%%%%
%Fig.3
%%%%%%%%%%%%%%%%%%%%%%%%%%%%%%%%%%%%%%%%%%%%%%%%%%%%%%%%%%
\begin{figure}[!ht]
	\centering\includegraphics[width=9cm]{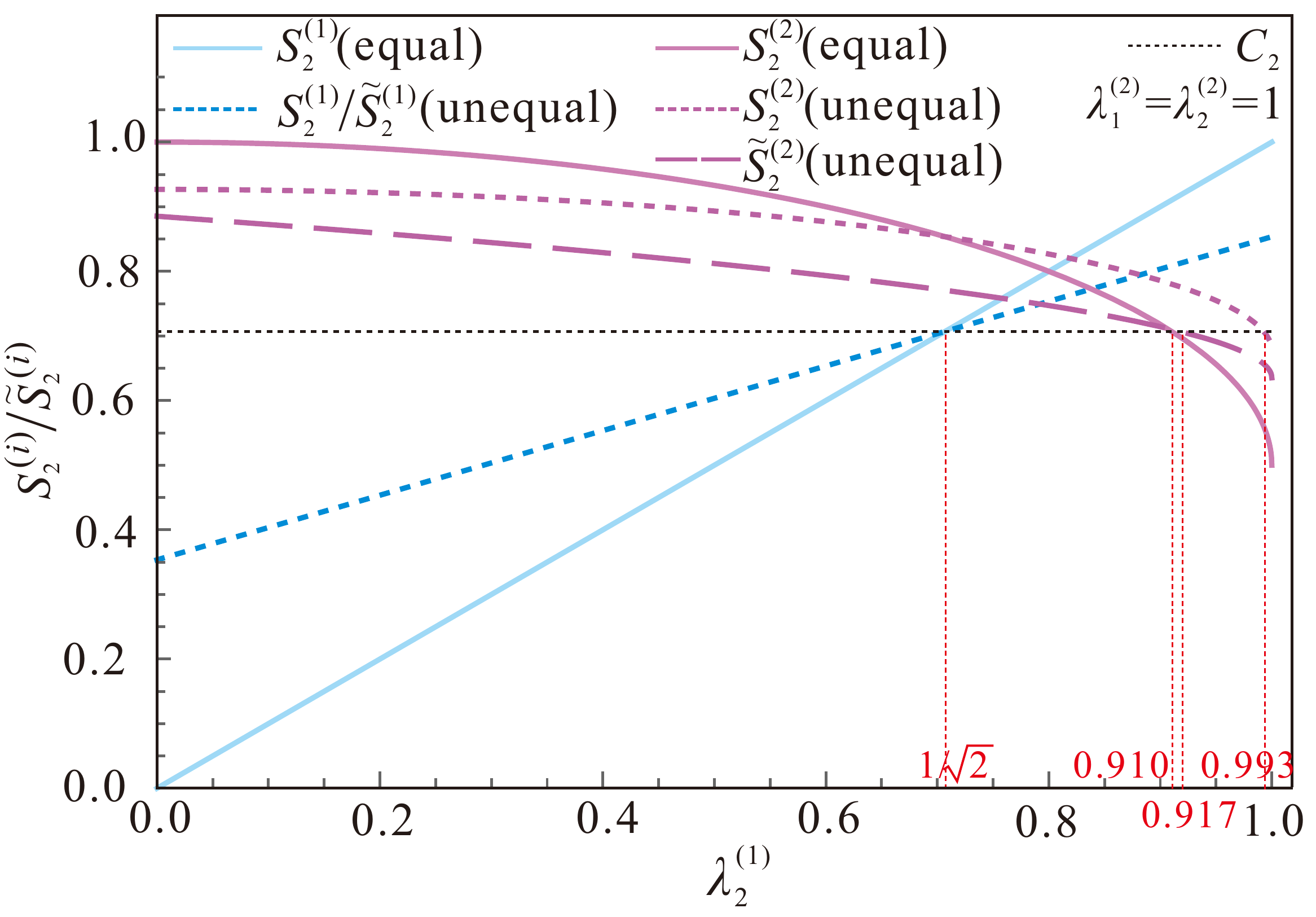}
	\caption{The steering parameters $S_2^{(i)}$ and $\widetilde S_2^{(i)}$ as functions of measurement strength $\lambda_2^{(1)}$. The solid line represents the situation when the measurement strengths are equal, while the dotted and dashed lines represent the situation when the measurement strengths are unequal. The blue and purple lines, respectively, indicate the steering parameters of $A_1$ and $B_1$, $A_2$ and $B_2$.}
	\label{FIG_3}
\end{figure}
%%%%%%%%%%%%%%%%%%%%%%%%%%%%%%%%%%%%%%%%%%%%%%%%%%%%%%%%%%
%Fig.4
%%%%%%%%%%%%%%%%%%%%%%%%%%%%%%%%%%%%%%%%%%%%%%%%%%%%%%%%%%
\begin{figure*}[!ht]
	\centering\includegraphics[width=18cm]{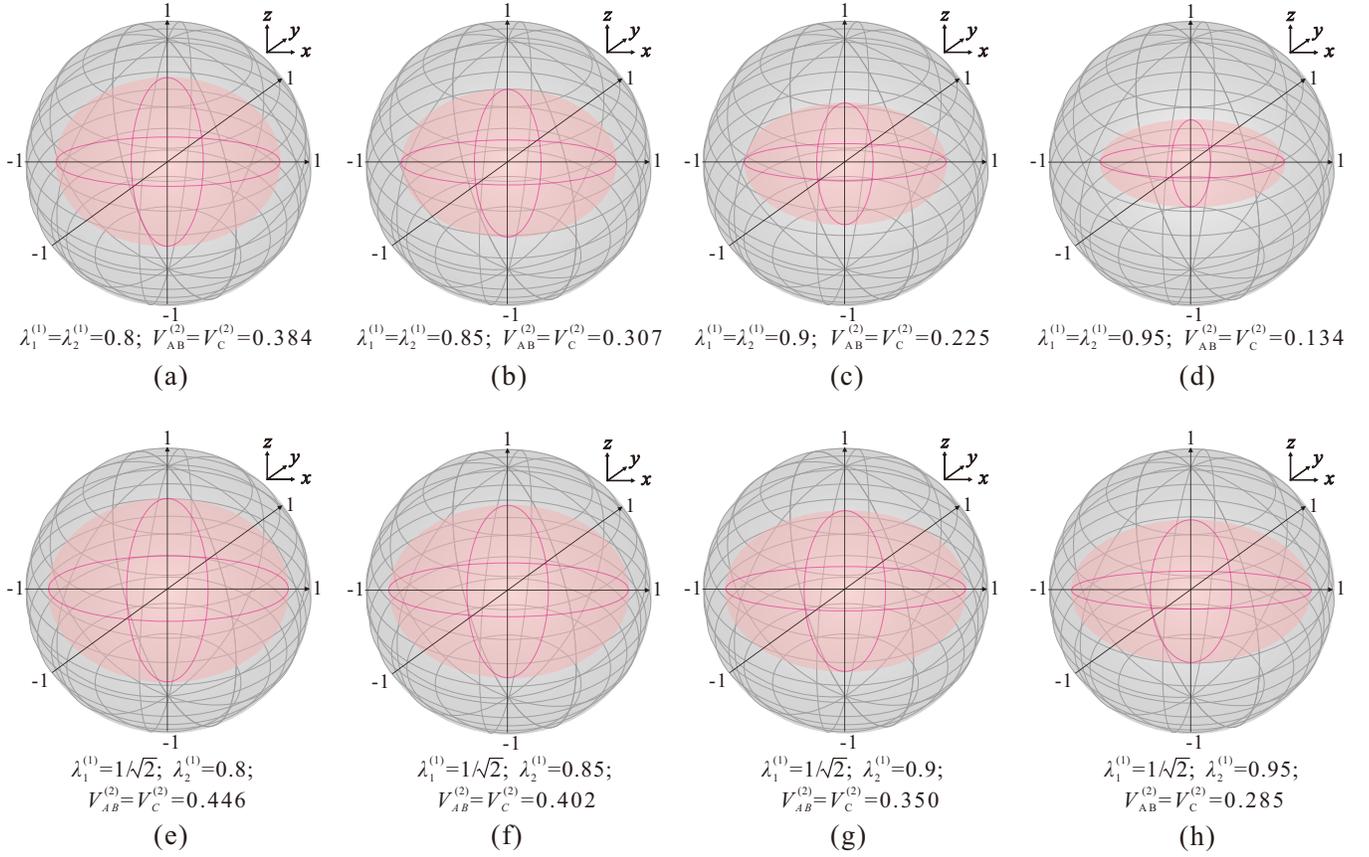}
	\caption{(a)-(d) The steering ellipsoids $\Omega_{AB}^{(2)}$ and $\Omega_{C}^{(2)}$ as a function of $\lambda_{1}^{(1)}$ and $\lambda_{2}^{(1)}$when $\lambda_{1}^{(1)}=\lambda_{2}^{(1)}$. (e)-(h) The steering ellipsoids $\Omega_{AB}^{(2)}$ and $\Omega_{C}^{(2)}$ as a function of $\lambda_{2}^{(1)}$ when $\lambda_{1}^{(1)}=1/\sqrt{2}$. The inner pink ellipsoids represent the steering ellipsoid, and the outer gray spheres represent the Bloch spheres.}
	\label{FIG_4}
\end{figure*}

\subsection*{The measurement settings with unequal strength}
Here, we relax the requirement that the strength of the two-setting nonlocal measurements used by each  pair of $A_i$ and $B_i$ need to be equal, i.e., $\lambda_{1}^{(i)}\neq\lambda_{2}^{(i)}$. We also relax the requirement that the strength of the two-setting local measurements used by each $A_i$ and $B_i$ need to be equal, i.e., $\eta_{1}^{(i)}\neq\eta_{2}^{(i)}$, $\gamma_{1}^{(i)}\neq\gamma_{2}^{(i)}$. However, we still set $\eta_k^{(i)} = \gamma_k^{(i)} = \sqrt{\lambda_k^{(i)}}$ is satisfied, where $k \in \{1, 2\}$. We find the maximum number of the pairs of $A_i$ and $B_i$ that can simultaneously share steering with Charlie can not be increased by using measurements with unequal strength. Figure~\ref{FIG_3} illustrates the effect of measurement strength $\lambda_{2}^{(1)} $ on the steering parameters in three cases: (i) using unequal strength local measurements (dotted blue line for $A_1$ and $B_1$, dashed purple line for $A_2$ and $B_2$), (ii) using equal strength nonlocal measurements (solid blue line for $A_1$ and $B_1$, solid purple line for $A_2$ and $B_2$) and (iii) using unequal strength nonlocal measurements (dotted blue line for $A_1$ and $B_1$, dotted purple line for $A_2$ and $B_2$). It is evident that the steering parameters of the first and second pairs of $A_i$ and $B_i$ can exceed the classical bound at the same time when the measurement strength $\lambda_{2}^{(1)} $ is increased to $1/\sqrt{2}$ in all three cases. Additionally, the range of $\lambda_{2}^{(1)}$ that steering can be shared simultaneously among the first and second pair of $A_1$, $B_1$, $A_2$, $B_2$ and Charlie is $(1/\sqrt{2},0.917)$ and $(1/\sqrt{2},0.910)$ in case (i) and case (ii) respectively, which can be further extended to $ (1/\sqrt{2},0.993)$ in case (iii). The results show that when $\lambda_{2}^{(1)}$ is in the range of $(0.917,0.993)$, the sharing of quantum steering can be further activated by using nonlocal measurements with unequal strength.

In addition, to provide a more intuitive visualization of the distinction between steering sharing activation of equal and unequal strength nonlocal measurements, we also examined how the steering ellipsoids $\Omega_{AB}^{(2)}$ of $A_2$ and $B_2$, as well as $\Omega_{C}^{(2)}$ of Charlie change as the measurement strengths $\lambda_{1}^{(1)}$ and $\lambda_{2}^{(1)}$ vary. 
The volume of the Charlie's steering ellipsoids generated by the measurements of $A_i$ and $B_i$ can be written as \cite{zhang2019experimental}
\begin{equation}
	V_{C}^{(i)}=\frac{|\det (T-{\widetilde{m}}{\vec{n}}^\top)|/(1-|{\widetilde{m}}{\vec{n}}|^2)^2}{4\pi/3}.
\end{equation}
Obviously, the volume of steering ellipsoids is normalized relative to the total volume of the Bloch sphere, which is $4\pi/3$ \cite{zhang2019experimental}. Similarly, the volume of the ellipsoid of $A_i$ and $B_i$ ($V_{AB}^{(i)}$), as generated by Charlie's measurements, can also be obtained. For the post-measurement state, the volumes $V_{AB}^{(2)}$ and $V_{C}^{(2)}$ are the same whether equal or unequal strength measurements are used because the steering ellipsoids $\Omega_{AB}^{(2)}$ and $\Omega_{C}^{(2)}$ are identical. The results of nonlocal measurements with equal strength are presented in the first row of Figure~\ref{FIG_4}, while those of nonlocal measurements with unequal strength are shown in the second row of Figure~\ref{FIG_4}. Clearly, with the increasing of $\lambda_{1}^{(1)}$ and $\lambda_{2}^{(1)}$, the steering ellipsoid contracts towards the sphere's center along all three principal axes simultaneously using equal strength measurements. However, when fixing $\lambda_{1}^{(1)}$ at $ 1/\sqrt{2} $, it can be observed that with the increase of $\lambda_{1}^{(2)}$, the steering ellipsoid only contracts towards the center of the sphere along the $y$ and $z$ axes. The length of principal axe in the $x$ direction is 0.854, remains constant. The principal axe shrinks in length in the $y$ direction at the same rate as in the case of equal strength measurements. 
In the $z$ direction, the principal axe shrinks in length at a rate that is approximately half as slow as it is when measured with equal strength. 
However, the volume reduction rate of the steering ellipsoid is usually slower when nonlocal measurements of unequal strength are employed. As the volume of the steering ellipsoid characterizes the steering ability, nonlocal measurements with unequal strength are more beneficial for activating quantum steering.

%%%%%%%%%%%%%%%%%%%%%%%%%%%%%%%%%%%%%%%%%%%%%%%%%%%%%%%%%%%%%%%%%%%
\section*{Discussion and conclusion}
To summarize, we have presented a scheme to activate quantum steering sharing among multiple pairs of $A_i$ and $B_i$ as well as a single Charlie using unsharp nonlocal measurements. Interestingly, we have found that unsharp nonlocal product measurements can be used to discover more steering sharing than their local counterparts, although they behave the same in terms of measurement outcome. Additionally, we have demonstrated that the steering sharing activation of nonlocal measurements can be further enhanced by replacing equal strength measurements with unequal strength measurements.
Compared to previous steering activation methods, our activation method has the advantage of not requiring the preparation of multicopy of the state, and does not weaken other types of correlations in the quantum system. This makes our method more experiment-friendly and resource-saving.

There are several relevant open problems that still need to be addressed. Firstly, it would be interesting to optimize the nonlocal measurement strategy to increase the number of observers who can share steering beyond the two pairs of sequential observers achieved in our work. This may be possible to achieve by employing unsharp nonlocal measurements on both sides, allowing sequential observers to share classical information and use an adaptive strategy, or by adopting mutually biased measurements, which requires further investigation in the future. Secondly, our method could potentially be applied to other types of quantum correlations, such as Bell nonlocality, quantum entanglement, quantum coherence, and quantum contextuality. This may promote the development of general relevant information protocols, such as quantum random access code, self-testing, and quantum randomness expansion.

\section*{Acknowledgements}

This work was supported by the National Natural Science Foundation Regional Innovation and Development Joint Fund (Grant No. U19A2075), the National Natural Science Foundation of China (Grant No. 12004358), the Fundamental Research Funds for the Central Universities (Grants No. 202041012 and No. 841912027), the Natural Science Foundation of Shandong Province of China (Grant No. ZR2021ZD19), and the Young Talents Project at Ocean University of China (Grant No. 861901013107).

\section*{Author contributions statement}

X. -H. H. designed theoretical simulation calculations and prepared the manuscript. T. Q. improved the manuscript. S. -C, D. tested the theoretical simulation results.  X. Y. improved the manuscript and supervised the research. Y. -J. G. planned, organized and supervised the project. All authors discussed the results and reviewed the manuscript. 

\section*{Additional information}

\textbf{Correspondence} and requests for materials should be addressed to Y. X. or Y. -J. G..

\textbf{Competing interests} The authors declare no competing interests. 

\textbf{Publisher’s note}: Springer Nature remains neutral with regard to jurisdictional claims in published maps and 
institutional afliations.

\end{document}